# Title: Floating under a levitating liquid

**Authors:** Benjamin Apffel[1,†], Filip Novkoski[1,†], Antonin Eddi[2], Emmanuel Fort[1,*]

**Affiliations:**

[1] ESPCI Paris, PSL University, CNRS, Institut Langevin, 1 rue Jussieu, F-75005 Paris, France.

[2] PMMH, CNRS, ESPCI Paris, Université PSL, Sorbonne Université, Université de Paris, F-75005, Paris, France

*Correspondence to: emmanuel.fort@espci.fr

†equal contributions.

**Abstract:** When placed upside down a liquid surface is known to destabilize above a certain size. However, vertical shaking can have a dynamical stabilizing effect. These oscillations can also make air bubbles sink in the liquid when created below a given depth. Here, we use these effects to levitate large volumes of liquid above an air layer. The loaded air layer acts as a spring-mass oscillator which resonantly amplifies the shaking amplitude of the bath. We achieve stabilization of half a liter of liquid with up to 20 cm width. We further show that the dynamic stabilization creates a symmetric Archimedes' principle on the lower interface as if gravity was inverted. Hence, immersed bodies can float upside down under the levitated liquid.

Maintaining a liquid upside-down is challenging as it inevitably tends to fall or trickle. However, various situations for which the liquid can be sustained are known. In the case of a limited surface size, capillary forces have a stabilizing effect (*1–3*). Alternatively, for large but thin liquid layers suspended under a plate, the wetting counterbalances the gravity (*4, 5*). In the latter case, the liquid interface does not stay flat but is destabilized in a regular pattern of hanging droplets. This instability driven by gravity is known as the Rayleigh-Taylor instability (*6, 7*). It occurs at the interface between two fluids whenever a denser one is placed over a lighter one. Several approaches have been used to stabilize the liquid layer such as temperature gradients (*8*), electric (*9*) or magnetic fields (*10*), rotational motion (*11*) and vertical vibrations (*12–14*). In the latter case, the amplitude of the vibration needs to be increased with the surface size. The maximum amplitude is set by the triggering of another instability called the Faraday instability which tends to destabilize fluid surfaces above a certain acceleration threshold (*15, 16*). However, this threshold can be raised by increasing the fluid viscosity (*17*). Hence, the upside-down liquid volume can be large provided the viscosity is properly chosen.

The vertical vibration of a fluid induces also other unexpected behavior when air bubbles are introduced in the fluid. Below a certain depth in the liquid, the bubbles are observed to sink defying the well-known Archimedes' principle (*18–20*). This effect has been studied for industrial applications in gas holdup and mixing in bubble column reactors (*21*).

Here, we study the effect of vibrations on the buoyancy of bodies immersed in levitating liquid layers. We first show that arbitrary large volume of liquids with surfaces larger than 10x10 cm$^2$ and half liter volume can be sustained on an air layer. The latter is produced by injecting more air into sinking bubbles to entirely fill the container surface. We then demonstrate that the vibrations endow the liquid with an upside-down buoyant (or Archimedes) force which allows objects to float under the liquid layer.

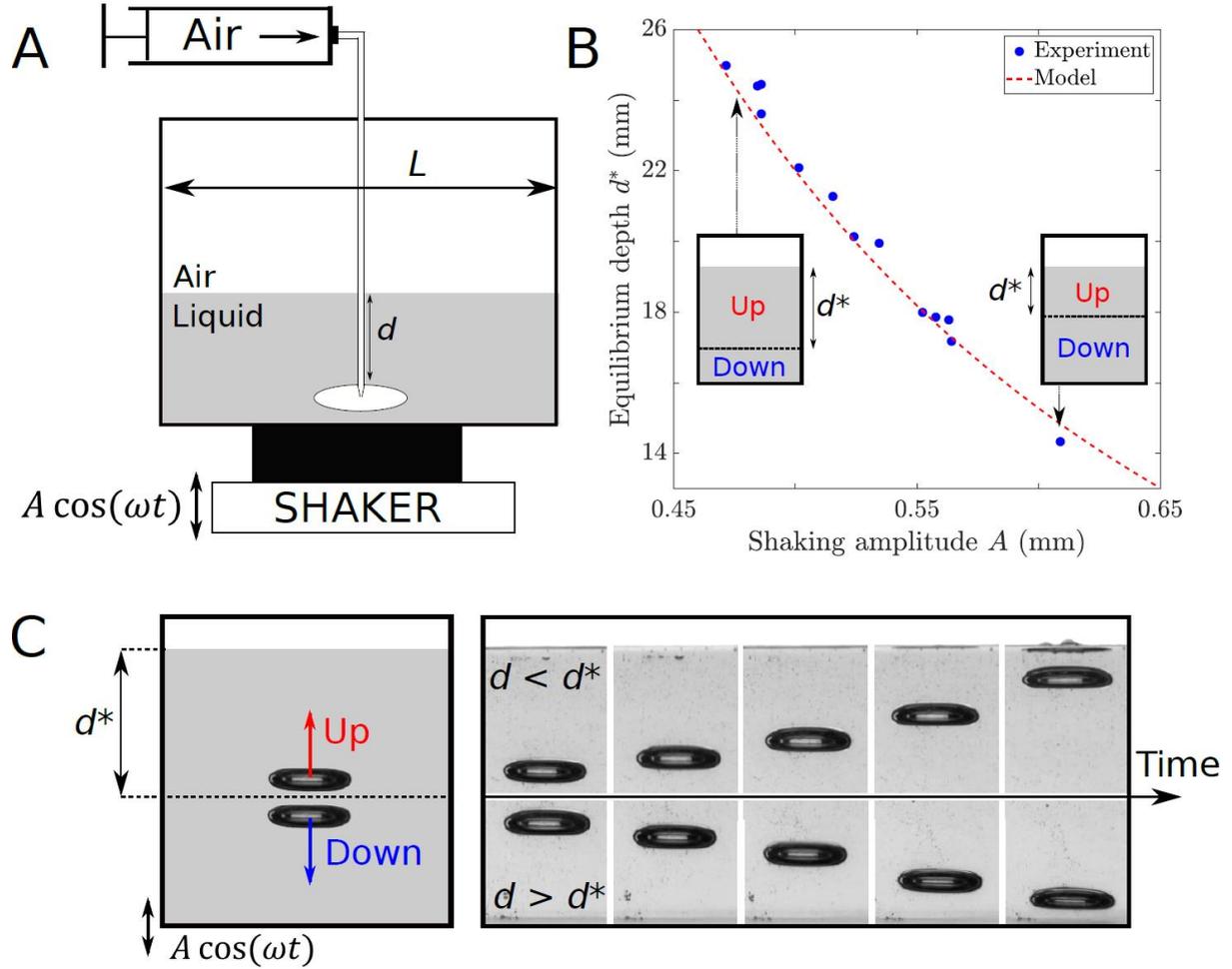

**Figure 1: Sinking bubbles in a shaken liquid bath.** (A) Experimental setup composed of a Plexiglas container of various sizes (up to 20 cm in width) attached on a vertically oscillating shaker with amplitude $A$ and frequency $\omega/2\pi$. The liquid is either glycerol or silicon oil with high viscosity (typically 0.5 Pa.s). The bubbles are created by injecting air with a syringe through a long needle immersed at depth $d$ into the liquid. (B) Evolution of the critical sinking depth $d^*$ as a function of the shaking amplitude $A$. Data are fitted by an invert quadratic scaling $c/A^2$ with $c = 5.5$ mm$^3$. (C) Schematics and image sequence of an ascending and descending bubble produced respectively above and below the critical sinking depth $d^*$ (see Supplementary Movie S1).

A Plexiglas tank fixed on a shaker is vibrated vertically at frequency $\omega/2\pi$ with an amplitude $A$ (Fig. 1A). The tank is filled with silicon oil or glycerol with high viscosity (typically ranging from 0.2 Pa.s to 1 Pa.s) to increase Faraday instability threshold (*13*). Air bubbles are created at various depths $d$ using a long needle connected to a syringe. Bubbles are observed to sink when placed below a critical depth $d^*$, while they reach the upper surface if created above $d^*$

(see snapshots Fig. 1C and Supplementary Movie S1). This behavior which defies standard buoyancy can be understood by a simple model taking into account the kinetic force exerted on the bubble in the oscillating bath (*18*, *20*). In a still bath, the buoyant force results from the difference in pressure between the top and bottom of the bubbles. This creates a constant upward force given by $F = \rho_l V g$ with $g$ the gravity acceleration, $V$ the volume of the body and $\rho_l$ the density of the fluid. In the oscillating bath, the effective gravity is modulated following $g_{\text{eff}}(t) = g + \gamma_{\text{exc}}(t)$ with $\gamma_{\text{exc}}(t) = -A\omega^2 \cos(\omega t)$ the bath acceleration. In addition, in the case of an air bubble at depth $d$, the pressure modulation induces a volume modulation $V(t)$ which can be considered isothermal and quasi-static for the low frequencies used (typically 100 Hz). The volume oscillations satisfy $V(t) = V_0 P_0/P(t)$ with $P(t)$ the gas pressure satisfying $P(t) = P_{\text{atm}} + \rho_l g_{\text{eff}}(t) h$ with $P_{\text{atm}}$ being the atmospheric pressure and the index 0 standing for the values at rest. When one averages the force $F = \rho_l V(t) g_{\text{eff}}(t)$ over one oscillation period, an additional sinking force, (also called the Bjerknes force (*22*)) arises. Below a critical depth $d_{\text{th}}^* \approx 2g P_0/(\rho_l A^2 \omega^4)$ corresponding to the equilibrium, the bubbles sink (*18*–*20*). Figure 1B shows $d^*$ as a function of the shaking amplitude $A$ in good agreement with an adjusted inverted quadratic scaling $1/A^2$. The very existence of such a depth results from the $\pi$-shift between the volume oscillations and the effective gravity ones.

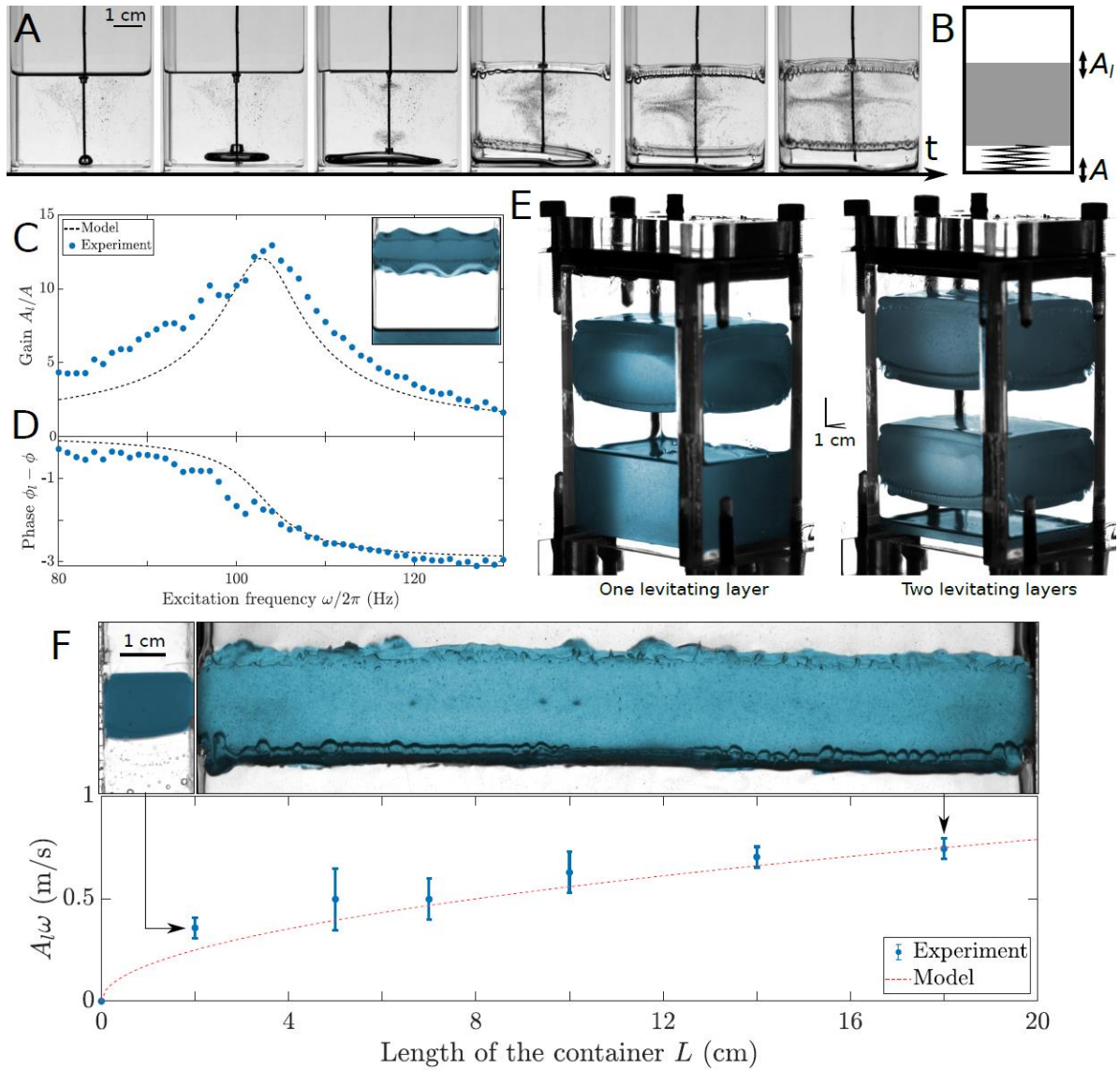

**Figure 2: Levitating liquid layer stabilized by Kapitza effect.** (A) Image sequence of the creation of the air layer obtained by blowing air at the bottom of the oscillating liquid bath through a needle. The sinking bubble grows until it completely fills the bottom of the bath (see Supplementary Movie S2). (B) Schematics of the spring-mass systems composed of the air layer loaded with the levitating liquid. (C) Enhancement of the liquid layer vertical amplitude $A_l/A$ and (D) relative phase shift $\phi_l - \phi$ of the liquid oscillations compared with that of the shaker as a function of the excitation frequency $\omega/2\pi$. Inset: image of Faraday instability triggered on the two opposite surfaces of the levitating liquid layer of silicon oil (see Supplement Movie S3). The experimental data (full circles) are fitted with the mass-spring model with fitting parameters $\omega/2\pi = 103$ Hz and $\Gamma = 0.04$ (dashed line, see Supplementary Text for details). (E) Colorized three-quarter views of the oscillating containers with one and two levitating liquid layers of silicon oil (see Supplementary Movies S4). (F) Threshold excitation velocity $A_l\omega$ for Kapitza stabilization of the liquid layer as a function of the length $L$ of the container: experimental data (circles) and model $A_l\omega = \sqrt{gL/\pi}$ (dashed line). Side views of the levitating bath in a 2 cm and 18 cm wide container are presented below (see Supplementary Movie S5).

The bubbles can be expanded by injecting air to completely fill the surface of the container, creating an air layer trapped below a levitating liquid layer (see the time sequence of Fig. 2A and Supplementary Movie S2) (*18*). The lower interface is stabilized by the vertical shaking preventing the release of the trapped air layer. This air layer is acting as a vertical spring loaded with the liquid mass placed upon it and driven by the shaker (Fig. 2B). It can be modeled as a driven damped harmonic oscillator $\ddot{z} + 2\Gamma\omega_{res}\dot{z} + \omega_{res}^2 z = A\omega^2\cos(\omega t)$ with $\omega_{res}$ the resonance frequency of the air layer and $\Gamma$ the damping ratio due to the shearing induced by the relative motion between the levitating liquid layer and the bath walls (see Supplementary Text). In the laboratory frame, the normalized oscillation amplitude $A_l(\omega)/A$ and its associated relative phase $\phi_l(\omega) - \phi$ compared to the shaker clearly show the expected resonance behavior (Fig 2C and 2D). The air layer thus enables the enhancement of the excitation amplitude of the shaker by more than one order of magnitude. Near the resonance, the amplitude is high enough to excite the Faraday instability on both sides of the fluid layer (see inset Fig. 2C and Supplementary Movie S3). This resulting "rain" emitted from the lower interface induces a thinning of the fluid layer which can be avoided by simply reducing the excitation amplitude. Provided the spring-mass oscillation is properly tuned, there is no restriction in the number of sustained levitating layers which can be piled up (see Fig. 2E and Supplementary Movie S4).

The vertical vibrations have a stabilizing effect on the lower fluid interface. This can be interpreted as a Kapitza effect which consists of a dynamical stabilization of an inverted pendulum by vertical shaking (*23*, *24*). Solving the Bernoulli equation for the fluid shows that the interface height $\zeta(k)$ at the spatial wavenumber $k$ behaves as an inverted pendulum. The spatial mode satisfies $\ddot{\zeta} + \left[\omega_0(k)^2 + \frac{A_l^2 k^2}{2}\omega^2\right]\zeta = 0$ with $\omega_0(k)^2 = -gk + (\gamma k^3)/\rho_l$ is the gravito-capillary dispersion relation with inverted gravity. Without vibrations, the oscillator is unstable for small enough $k$ ($\omega_0(k)^2 < 0$) leading to the Rayleigh-Taylor instability while large wave numbers are stabilized by capillarity. The last term in the equation arises from the modulation of the effective gravity. In gravitational regime, the stabilization is reached for wavenumbers satisfying $k > 2g/A_l^2\omega^2$ (see Supplementary Text). The limited size $L$ for the bath sets a limit to the observed excitable wavenumber $k > 2\pi/L$ (only anti-symmetric modes satisfying volume conservation are considered). The stability if the interface stability is thus obtained for oscillating velocities $A_l\omega > \sqrt{gL/\pi}$. Although there seems to be no size limit for stabilization, viscosity must always be high enough to prevent the Faraday instability to occur (see inset Fig. 2C). Figure 2E shows the oscillating velocity $A_l\omega$ needed to stabilize baths with lengths $L$ up to 18 cm (insets show views of levitating layers for $L = 2$ cm and $L = 18$ cm, see Supplementary Movie S5).

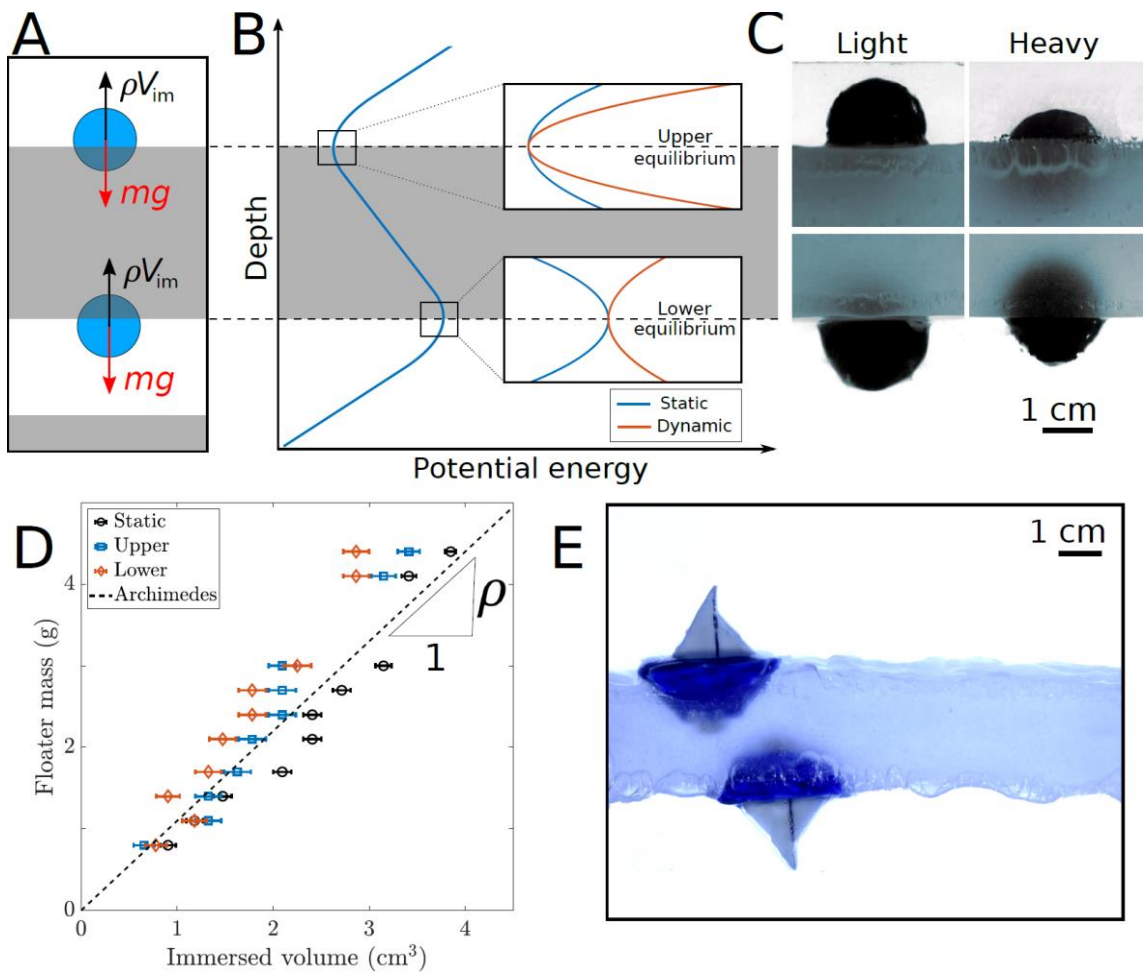

**Figure 3: Archimedes' principle over and under a levitating liquid layer.** (A) Schematics of the force balance at the two opposed interfaces with buoyant force cancelling the weight of immersed bodies. (B) Typical profile of the static potential along the vertical direction $z$ neglecting the dynamical effects. Two equilibrium positions appear at each interface, the lower one being unstable. Insets: close up of the potential near the equilibrium positions with the addition of the dynamical stabilizing effect (red

line, see Supplementary Text). (C) Side views of 2 cm diameter plastic spheres floating upwards and downwards with lower (left) and higher density (right). (D) Equilibrium positions for 2 cm diameter spheres with various masses as a function of the immersed volume at the upper (squares) and lower interface (diamonds). Circles are equilibrium positions obtained without shaking. The dashed line is given by the Archimedes' principle with $\rho_l = 1.1$ kg/L for glycerol. (F) Boats floating over and under a levitated liquid layer (see Supplementary Movie S6).

We now focus on the possibility of having objects floating at the inverted interface of the levitating fluid layer, i.e. upside-down floating. Archimedes' principle states that the upward buoyant force exerted on an immersed body, whether fully or partially submerged, is equal to the weight of the displaced fluid. Although this may seem counterintuitive, the transpose symmetric position at the lower interface (see Fig 3A) also exhibits an upward buoyant force equal to the weight of displaced liquid. Figure 3B shows the typical potential exerted on a floating body without taking into account the dynamic effects (see supplementary materials for details). The two equilibrium positions associated with each interface are clearly visible. However, while the upper position is stable, the lower is not: pushing the body down would make it fall and up would make it float on the upright interface. Taking into account the dynamical effect, i.e. the time averaged effect of the oscillations, provides an additional stabilizing dynamical potential around the two equilibrium positions (see inset Fig. 3B and Supplementary Text). This leads to an extra stabilization of the upper equilibrium position and more spectacularly to the stabilization of the lower one provided high enough accelerations. It is thus possible to have floating bodies with varying density above and under the levitating liquid layers (see Figure 3C). The equilibrium positions satisfy the standard Archimedes' principle on both interfaces, with symmetric equilibrium positions (Fig. 3D). Hence the vibration not only gives stability of the lower horizontal interface of a liquid but also permits a vertical stabilization of the unstable equilibrium position that a floater would experience on such interface. This dynamical "anti-gravity" enables boats to float on both interfaces (Fig. E, see Supplementary Movie S6) and even to sink upwards.

**Acknowledgments:** The authors are grateful to Suzie Protière, Arnaud Lazarus, Sander Wildeman and the colleagues and students of "Projets Scientifiques en Equipes" for insightful discussions.

**Funding:** The authors thank the support of AXA research fund and the French National Research Agency LABEX WIFI (ANR-10-LABX-24).

**Supplementary Text**

Spring-mass model for the levitating liquid layer

We consider a liquid layer maintained above a layer of air by shaking vertically. In the laboratory frame, the position of the bottom interface of the air layer moves with the container and satisfies $z_b(t) = A\cos(\omega t + \phi)$ with $A$, $\omega$ and $\phi$ being the amplitude, the angular frequency and the phase of the forcing. The position of the upper interface of the air layer is given by $z_l(t)$. The height of the air layer is thus given by $z_a(t) = z_l(t) - z_b(t)$. This corresponds also to the relative motion of the upper interface, and thus of the levitating liquid layer relative to the container (see Figure S1).

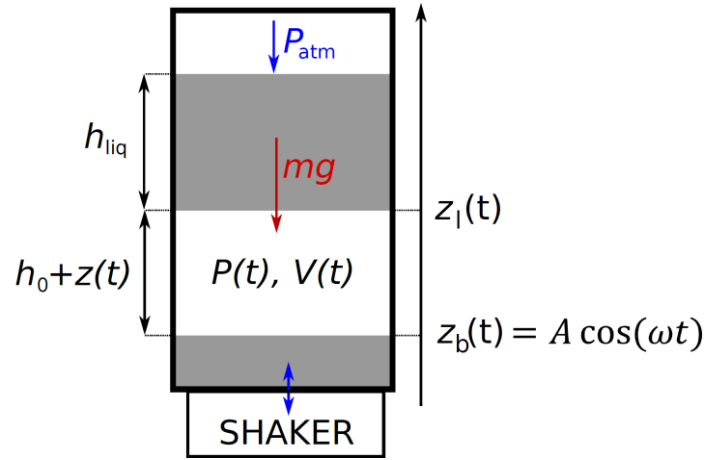

**Fig. S1.** Sketch of a levitating liquid layer with notations

The air layer is considered as a perfect gas and its pressure $P(t)$ and volume $V(t)$ are assumed to verify:

$$P(t)V(t)^\beta = P_0 V_0^\beta \qquad (1)$$

with $\beta$ being the polytropic index, $P_0$ and $V_0$ the static pressure and volume respectively. In addition, $P_0 = P_{atm} + mg/S$, $m$ being the mass of the levitating liquid layer and $V_0 = Sh_0$, $h_0$ being the height of the air layer at rest. Depending on the transformation the value of $\beta$ can vary: $\beta = 1$ for an isothermal process while $\beta$ is equal to the heat capacity ratio for an adiabatic one.

For small oscillations observed experimentally around the mean height $h_0$, we can perform a Taylor expansion of eq. (1) and get the pressure in the gas as

$$P(t) \approx P_0(1 - \beta z(t)/h_0) \qquad (2)$$

With $z(t) = z_a(t) - h_0 \ll h_0$ for small oscillations. Writing the Newton's law for the liquid layer, we obtain

$$m\ddot{z}_l = -mg - SP_{atm} + SP_0\left(1 - \frac{\beta z}{h_0}\right) - c\dot{z} \qquad (3)$$

with $c$ coming from damping.

On the right hand side of the equation, the first term is the weight of the liquid layer, the second term results from the air pressure on both sides of the liquid layer and the third term is damping due to the relative motion between the container and the liquid.

This can be rewritten as a damped harmonic oscillator with a sinusoidal forcing (as in the paper)

$$\ddot{z} + 2\Gamma\omega_{res}\dot{z} + \omega_{res}^2 z = -\ddot{z}_b \tag{5}$$

with the resonant frequency satisfying $\omega_{res}^2 = \beta P_0 S/mh_0$ and the damping ratio $\Gamma = \frac{c}{2m\omega_{res}}$. Hence the motion of the liquid in the laboratory frame writes $z_l(t) = A_l \cos(\omega t + \phi_l)$.

We finally have, with $\widehat{\omega} = \omega/\omega_{res}$, for the relative amplitude

$$\frac{A_l}{A} = \sqrt{\frac{1+4\Gamma^2\widehat{\omega}^2}{(1-\widehat{\omega}^2)^2+4\Gamma^2\widehat{\omega}^2}} \tag{6}$$

and for the relative phase

$$\phi_l - \phi = \operatorname{atan} 2\Gamma\widehat{\omega} - \operatorname{atan} \frac{2\Gamma\widehat{\omega}}{1-\widehat{\omega}^2} \tag{7}$$

These expressions were used to produce the fits in Figure 2C and 2D respectively. We obtain for the damping ratio $\Gamma \approx 0.04$ and for $\omega_{res}/2\pi \approx 103$ Hz. The experimental conditions gives $h_0 \approx 1.5$ cm and the liquid layer height $h_l \approx 2.5$ cm. Using these values and the glycerol density $\rho = 1.1$ kg/L, we can estimate $\beta \approx 1.7$. This value is slightly higher than expected, one would rather expect a value in the range $1 \leq \beta \leq 1.4$ between an isothermal transformation to an adiabatic one.

Stabilization of an inverted liquid/air interface by vertical oscillations

We focus our attention on the evolution of small perturbations at the interface of two incompressible and immiscible fluids. $\rho_{up}$ and $\rho_{down}$ are the densities of the upper and lower fluids respectively. The interface is placed at $z = 0$ and the perturbations at position $\boldsymbol{r}$ and time $t$ in the horizontal plane are $\zeta(\boldsymbol{r},t)$.

It is convenient to study the interface deformations in the spatial Fourier space, $\zeta(\boldsymbol{k})$ is the deformation at the spatial frequency $k = |\boldsymbol{k}|$ and in the direction $\boldsymbol{k}$.

Assuming the fluid velocities at the interface are small, it is possible to linearize Bernoulli's equations and show that each interface deformation $\zeta(\boldsymbol{k})$ behaves as a harmonic oscillator (24)

$$\ddot{\zeta}(\boldsymbol{k}) + \omega_0(k)^2 \zeta(\boldsymbol{k}) = 0 \tag{8}$$

with the angular frequency $\omega_0(k)$ satisfying the dispersion relation $\omega_0(k)^2 = -Agk + \gamma k^3/(\rho_{up} + \rho_{down})$, $g$ the gravitational acceleration, $\gamma$ the surface tension and $A = (\rho_{up} - \rho_{down})/(\rho_{up} + \rho_{down})$ is the Atwood number.

We now consider the standard configuration with a liquid of density $\rho_l$ and the air interface above. Neglecting the density of air, $\omega_0(k)^2 = gk + \gamma k^3/\rho_l$ which is the standard gravity-capillary dispersion relation. In this case, both gravity effects and capillary ones are positive and thus stabilize the interface resulting in the propagation of waves along the interface.

We now assume that the liquid stands above the air i.e. we turn the previous interface upside down. In this case the dispersion relation writes: $\omega_0(k)^2 = -gk + \gamma k^3/\rho_l$. The effect of the gravity on the interface is reversed while capillarity is not affected. The gravity term is negative inducing a destabilization of the interface.

For small wave vectors for which gravity effects dominate, $\omega_0(k)^2 < 0$ so that a perturbation of the interface will exponentially increase in time and the liquid will eventually fall down. We recover here the well-known Rayleigh-Taylor instability that occurs when a heavier fluid is set above a lighter one. The most unstable mode occurs for the minimum (negative) value of $\omega_0(k)^2$ which is associated to the wavenumber $k = \sqrt{\rho g/2\gamma}$. For large wave vectors, capillarity dominates and it has a stabilizing effect.

We are now focusing on dynamical stabilization induced by vertical vibrations of the fluid.

We assume that the liquid is submitted to a vertical oscillation $A_l \cos(\omega t)$. The acceleration of the liquid is then $-A_l \omega^2 \cos(\omega t)$. The acceleration of the liquid can be interpreted as a modulation of the effective gravity $g_{eff}(t) = g - A_l \omega^2 \cos(\omega t)$ in the accelerated frame.

Eq. (8) can thus be modified using $g_{eff}$ and we obtain

$$\ddot{\zeta} + \omega_0(k)^2 \zeta = -kA_l\omega^2 \cos(\omega t)\, \zeta \tag{9}$$

We now perform an analysis similar to the one used by Kapitza for the stabilization of the inverted pendulum shaken vertically (23). We decompose $\zeta$ into a slow component $\zeta_s$ and a fast one $\zeta_f$ oscillating at $\omega$ so that $\zeta = \zeta_s + \zeta_f$. This decomposition is justified for $k$ modes satisfying $|\omega_0(k)| \ll \omega$. In this case, the evolution of the height $\zeta(k)$ is slow compared to the modulation. For all unstable modes, this hypothesis is well verified in our experimental conditions.

We can then expect that the vibration will induce a slow additional stabilizing effect on $\zeta_s$ induced by a fast perturbation $\zeta_f$ at angular frequency $\omega$. We denote with $<.>$ the mean over one oscillation period. We have $<\zeta_f> = 0$ and $<\zeta_s \cos(\omega t)> = 0$. Writing the equation of motion and keeping only terms oscillating at $\omega$ and first order term in $\zeta_f$ gives $\ddot{\zeta}_f \approx -kA_l\omega^2 \cos(\omega t)\zeta_s$ so that $\zeta_f \approx kA_l \cos(\omega t)\zeta_s$. We now take the mean $<.>$ of the evolution eq. (9) and get:

$$\ddot{\zeta}_s + \omega_0(k)^2 \zeta_s = -kA_l\omega^2 < \cos(\omega t)\zeta_f > = -\frac{k^2 A_l^2}{2}\omega^2 \zeta_s \tag{10}$$

Or equivalently

$$\ddot{\zeta} + \left[\omega_0(k)^2 + \frac{A_l^2 k^2}{2}\omega^2\right]\zeta = 0 \qquad (11)$$

We recover here the equation given in the main text and find that the new dynamical term resulting from the vibration is positive and thus tends to stabilize the interface.

This equation is valid as long as $|\omega_0(k)| \ll \omega$. For large wave vectors, $|\omega_0(k)| \ll \omega$ does not hold anymore, however these modes are stabilized by capillarity ($\omega_0(k)^2 > 0$). For the unstable modes with small $k$ for which $\omega_0(k)^2 < 0$, the condition $|\omega_0(k)| \ll \omega$ can always be satisfied. The condition to stabilize a mode with a given $k$ is given by $k > \frac{2g}{A_l^2 \omega^2}$.

Hence, for an interface of size $L$, the excitable wavenumbers satisfy $k > \pi/L$, and one has to apply an excitation velocity $A_l\omega > \sqrt{2gL/\pi}$.

Note that in the experiments, the air layer trapped under the liquid tends to mute the anti-symmetric modes on the interface ($k = \pi q/L$ with $q$ being an odd number). These modes do not conserve the volume of the air layer. The stability of an interface of length $L$ in this case should satisfy $A_l\omega > \sqrt{gL/\pi}$ as measured experimentally (see Fig. 2F).

Dynamical stabilization of floating bodies at the lower and upper interface of a levitated oscillating liquid layer

a. Buoyant force in the rest frame

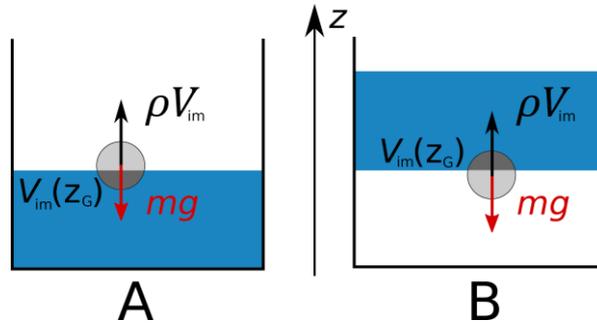

**Fig. S2.** Sketchs of the floating equilibrium on the upper interface (A) and on the inverted interface (B)

We consider a body of mass $m_b$, volume $V_b$ and density $\rho_b = m_b/V_b$ floating on a bath of liquid of higher density $\rho_l > \rho_b$. The capillary effects are neglected.

The buoyant force is due to hydrostatic pressure in the liquid, and it satisfies

$$\boldsymbol{F}_{\text{buoy}} = -\int_S P d\boldsymbol{S} = -\int_V \boldsymbol{\nabla} P dV = \rho_l g V_{\text{im}}(z_G) \qquad (12)$$

With $P$ the pressure exerted on the body surface $S$, $V$ is the body volume and $V_{\text{im}}(z_G)$ is the immersed volume with the center of mass at vertical position $z_G$ (see Figure S2A). We consider that the immersed volume is fully characterized by $z_G$ i.e. it does not depend on the orientation of the body and the center of pressure coincides with the center of mass, or it is determined for a given body orientation if the center of pressure is different from the center of mass. In the latter case, the stable orientation is obtained for the center of pressure positioned above the center of mass.

Newton's second law applied to the floating body for the position of its center of mass $z_G$ gives

$$m_b \ddot{z}_G = -m_b g + \rho_l g V_{im}(z_G) \tag{13}$$

At equilibrium, $V_{im}(z_{G,eq}) = m_b/\rho_l$ (14)

If we now consider small displacements $Z$ around this equilibrium position so that $z_G = z_{G,eq} + Z$, we obtain

$$\ddot{Z} + \omega_{buoy}^2 Z = 0 \quad \text{with} \quad \omega_{buoy}^2 = -\frac{\rho_l g}{m_b} \frac{dV_{im}}{dz}(z_{G,eq}) > 0 \tag{15}$$

The position is stable and the floater oscillates around its equilibrium position. For a sphere of diameter 2 cm and of relative density $0.2 \leq \rho_b/\rho_l \leq 0.8$, one has typically $1 \leq \omega_A \leq 10$ rad/s.

b. <u>Dynamical effects on floating bodies on the vibrated liquid interface</u>

We now focus on the effect of an added vertical acceleration of the liquid with angular frequency $\omega$ and amplitude $A$ on the equilibrium position. In this case the liquid is submitted to an oscillating effective gravity $g_{eff}(t) = g - A\omega^2 \cos(\omega t)$ with the proper choice of the time origin.

In the low excitation frequency range used in the experiment ($\omega/2\pi < 150$ Hz), the hydrostatic pressure in the liquid and the constant atmospheric pressure outside still holds. The buoyant force is thus given by eq. (12) using the effective gravity: $F_{buoy} = \rho_l g_{eff} V_{im}(z_G)$.

To write Newton's second law, we assume that the floating body moves with the liquid and that there is no relative motion between the body and the surrounding liquid. This hypothesis is verified by experimental observations. Under this assumption, the equation of motion in the accelerated frame is given by eq. (13) using the effective gravity

$$m_b \ddot{z}_G = -m_b g_{eff} + \rho_l g_{eff} V_{im}(z_G) \tag{16}$$

The equilibrium position is the same as for the static liquid and is given by eq. (14)
Following the previous analysis for small displacements $Z$ around the equilibrium position, we obtain

$$\ddot{Z} + \omega_{buoy}^2 Z = \omega_{buoy}^2 \frac{A\omega^2}{g} \cos(\omega t) Z \tag{17}$$

We now assume that $\omega_{buoy} \ll \omega$ which is always verified in our experimental cases and we use Kapitza's approach for the dynamical stabilization of an inverted pendulum using vertical shaking <u>(1, 2)</u>. We decompose $Z$ into a slow component $Z_s$ and a fast one $Z_f$ at the excitation frequency $\omega$ so that $Z = Z_s + Z_f$. Denoting $<.>$ the mean value over the excitation period, the two components satisfy $< Z_s \cos(\omega t) > = 0$ and $< Z_f > = 0$. Under these conditions, we find for the oscillating terms at $\omega$ at first order in $Z_f$ (3)

$$\ddot{Z}_f = \omega_{\text{buoy}}^2 \frac{A\omega^2}{g} \cos(\omega t) Z_s \tag{18}$$

Hence,

$$Z_f = -\omega_{\text{buoy}}^2 \frac{A}{g} \cos(\omega t) Z_s \tag{19}$$

Averaging eq. (17) over an excitation period and using eq. (19), we obtain

$$\ddot{Z}_s + \omega_{\text{buoy}}^2 Z_s = -\frac{\omega_{\text{buoy}}^4}{2} \left(\frac{A\omega}{g}\right)^2 Z_s \tag{20}$$

This gives

$$\ddot{Z}_s + \omega_{\text{buoy}}^2 (1 + \alpha) Z_s = 0 \quad \text{with } \alpha = \frac{\omega_{\text{buoy}}^2}{2} \left(\frac{A\omega}{g}\right)^2 \tag{21}$$

The term $\alpha$ is a dynamical term resulting from the non-zero correlation between the floating object motion and the modulated pressure due to the vibration. Since $\alpha > 0$, this term induces an additional stabilization of the equilibrium floating position (see Fig. 3B).

c. Dynamical stabilization of floating bodies on the lower interface of a levitating liquid

We consider the case of the lower interface in a levitating liquid layer (see Fig. S2B). The liquid layer is stabilized by vertical oscillations with an amplitude $A$ and an angular frequency $\omega$. We assume that such an inverted interface exists and is stable regardless of its underlying stabilization mechanism induced by the vibration.

We first focus on the existence of a static equilibrium at each interface. Under the hypothesis of a hydrostatic pressure in the liquid and a constant pressure in the trapped air layer, the static equation (13) still applies. Two equilibrium positions can thus be found for a floating body, both satisfying eq. (14). The standard equilibrium on the upper interface and another equilibrium on the lower interface (see Fig. 3B).

As an object is moved through the lower interface, its immersed volume $V_{\text{im}}$ varies from the entire body volume to zero (when entirely in the air layer). Hence, the symmetric equilibrium positions on the two interfaces satisfying $V_{\text{im}}(z_{G,\text{eq}}) = m_b/\rho_l$ are equilibrium positions.

However, for the lower interface, this position is not stable, pushing the body slightly upward, increases the immersed volume and results in a net upward force while pushing the body downward reduces the immersed volume and makes the body fall.

This originates from the positive sign of $\frac{dV_{\text{im}}}{dz} > 0$ for the lower interface which results in a negative sign for $\omega_{\text{buoy}}^2 = -\frac{\rho_l g}{m_b} \frac{dV_{\text{im}}}{dz}(z_{G,\text{eq}}) < 0$.

Taking into account the vertical oscillations and following the same approach as in the previous section leads to the same eq. (16). For small displacements $Z$ around the lower equilibrium position with a fast dynamics $Z_f$ and a slow one $Z_s$, one obtains

$$\ddot{Z}_s + \omega_{\text{buoy}}^2(1+\alpha)Z_s = 0 \quad \text{with} \quad \alpha = \frac{\omega_{\text{buoy}}^2}{2}\left(\frac{A\omega}{g}\right)^2 \tag{22}$$

However, now $\omega_{\text{buoy}}^2 < 0$ and thus $\alpha < 0$.

The equilibrium is stable if $\omega_{\text{buoy}}^2(1+\alpha) > 0$ which is possible if $\alpha < -1$.

One obtains the following stability condition for the amplitude of the forcing velocity

$$A\omega > \frac{\sqrt{2}g}{|\omega_{\text{buoy}}|} \tag{23}$$

Note that for objects with a non-coincident center of pressure and center of mass, the former should always be above the latter for stability. However, experimentally an equilibrium is easily found even in the opposite orientation of the body (see Fig. 3E).